\documentstyle[12pt,aaspp4]{article}

\begin{document}

\def\kms{km~s$^{-1}$}
\def\msun{$M_{\odot}$}
\def\rsun{$R_{\odot}$}
\def\lsun{$L_{\odot}$}
\def\halpha{H$\alpha$}
\def\Teff{T$_{\rm eff}$}
\def\logg{$log_g$}

\tighten

\title{ Near-Infrared Spectral Classification of late-M and L Dwarfs}

\author {I. Neill Reid}
\affil {Dept. of Physics \& Astronomy, University of Pennsylvania, 209 S. 33rd
Street, Philadelphia, PA 19104-6396;
e-mail: inr@herschel.physics.upenn.edu}

\author {A. J. Burgasser}
\affil {Dept. of Physics, 103-33, California Institute
of Technology, Pasadena, CA 91125}

\author {K. L. Cruz}
\affil {Dept. of Physics \& Astronomy, University of Pennsylvania, 209 S. 33rd
Street, Philadelphia, PA 19104-6396}

\author {J. Davy Kirkpatrick}
\affil {Infrared Processing and Analysis Center, 100-22, California Institute
of Technology, Pasadena, CA 91125}

\author {J. E. Gizis}
\affil {Department of Physics and
Astronomy, University of Massachusetts, Amherst, MA 01003}

\begin {abstract}

We present near-infrared (1 to 2.5 $\mu$m) low-resolution spectroscopy of
eleven ultracool dwarfs with spectral types ranging from M7 to L8. 
Combining our observations with data published previously by Leggett 
{\sl  et al.} (2001), we have measured equivalent widths for the strongest
atomic features and constructed narrowband indices to gauge the strength of
the strongest molecular features. Those measurements show that the behaviour
at near-infrared wavelengths is well correlated with spectral type, where the
latter is defined from observations between 6300 and 10000 \AA.
In particular, four indices designed to measure absorption in the wings of the 
1.4 and 1.85$\mu$m steam bands exhibit a linear, monotonic
correlation with spectral type on the Kirkpatrick
{\sl  et al.} (1999) system, allowing consistent calibration from both
optical and  near-infrared spectroscopy.

\end{abstract}


\section {Introduction}

Spectral classification has played a vital role in astronomical research for
almost 140 years. Originating with Secchi's (1866) division of the brightest stars
into Classes I (Sirius-like), II (Sun-like) and III (Betelgeuse-like), 
the search for pattern and order has been
crucial in understanding  stellar structure and evolution. The classic Harvard system
was developed by Fleming, Maury and Pickering in the 1890s, re-organised and refined by Cannon
in the 1900s, and codified by Morgan, Keenan \& Kellman
in the 1940s (Morgan {\sl et al.}, 1943). Boeshaar (1976) and
Kirkpatrick, Henry \& McCarthy (1991) extended calibration to
spectral type M9.5. Yet, for all these additions, the basic scheme remained the Harvard system,
running through OBAFGKM along the main sequence. That classification 
proved sufficient until the last few years of the twentieth century, 
when observations finally outstripped the system.

Spectral class M is characterised by the presence of TiO absorption. The advent of
deep, wide-field near-infrared sky surveys catalysed both the discovery of numerous
ultracool dwarfs with weak or no TiO absorption, and the creation of
two new spectral types: class L, prototype GD 165B 
(Becklin \& Zuckerman, 1988; Kirkpatrick, Henry \& Liebert, 1993), characterised by
metal-hydride bands and strong alkali lines; and class T, prototype Gl 229B
(Nakajima et al, 1995), characterised by methane absorption at 1.6 and 2.1 $\mu$m.
Extensive follow-up optical spectroscopy has led to the definition of sub-types, L0 to
L8, for the former class (Kirkpatrick et al, 1999 - hereinafter, K99). 
At present, classification rests on molecular bandstrengths in the far red, between 0.6
and 1.0 $\mu$m, and acquiring spectra of adequate signal-to-noise at those wavelengths is a 
time consuming process for even the brightest L dwarfs. This technical problem can
be circumvented to a large extent by calibrating the classification system at near-infrared
wavelengths, where these cool dwarfs are significantly brighter. That 
issue is addressed in the current paper.

We have obtained intermediate resolution 0.9 to 2.5$\mu$m spectra of a small sample of 
ultracool dwarfs with spectral types between M8 and L8. 
Our aim is the identification of
suitable spectral-type indicators at those wavelengths, combining our own spectra with
similar data from the literature, notably Leggett {\sl et al.'s} (2001 - hereinafter, L2001) 
recent observations. We use these observations to confirm that variations in 
the more prominent near-infrared  features preserve, by and large, the L0, L1...L8 
sequence defined from optical spectroscopy.  Rather than define an independent
infrared classication system, we provide calibrating relations, based on the
depth of the 1.4 and 1.85 $\mu$m steam bands, which transfer the
Kirkpatrick {\sl  et al.} far-red system to near-infrared wavelengths. 
Section 2 describes the  acquisition, reduction and calibration of our 
near-infrared spectra; section 3 discusses the photometric and spectroscopic properties
of late-M and L dwarfs, and identifies spectrophotometric indices suitable for spectral
classification; and section 4 summarises our conclusions.

\section {Observations}

\subsection {Data acquisition}

Our observations were obtained on 19-22 December 1998 (UT) using the cooled grating spectrograph CGS4,
mounted  on the UK Infrared Telescope. Some cirrus was present during parts of the last two nights of the
run, but seeing  conditions were 1 arcsecond or better throughout. We used a 1.2 arcsecond slit, matched
to 1 pixel on the detector, a $256 \times 256$ InSb array. 
We employed the 40 l/mm grating with the 150 mm camera to cover the near-infrared JHK r\'egime
in four settings: second order, centred at 1.10 and 1.35 $\mu$m, giving a  resolution of 18\AA\, 
or 550 kms$^{-1}$
(designated J1 and J2); and first order, centred at 1.75 and 2.25 $\mu$, at resolution 1,100 kms$^{-1}$
(designated H and K). The full spectral range covered by the four passbands is $\sim0.95$ to 2.6 $\mu$m,
although terrestial CO$_2$ absorption affects wavelengths beyond 2.5 $\mu$m. 

We observed eleven ultracool dwarfs, ranging in spectral type from M7 to L8.
Table 1 lists the targets and the journal of observations. The observations were made using a
$2 \times 2$ sampling grid to minimise the effect of bad pixels, 
with individual (single frame)
exposure times of 20 seconds at H and K and 30 seconds at
J1 and J2. Flat-field exposures were obtained immediately preceding each programme
star observation.
Primary wavelength calibration is provided by xenon and krypton arc lamps. 
Each observation consisted of a series of paired exposures, nodding the telescope to move
the star along the slit to permit accurate sky subtraction. The programme star
observations, filter by filter,  were interspersed with observations of bright F-type stars at small angular
separation. Those stars serve to calibrate both absorption due to the terrestial atmosphere and the
overall energy distribution. The relevant reference stars for each source are listed in Table 1.

\subsection {Data reduction}

The successive pair-subtracted observations  were combined using standard UKIRT
programs to give a single image in each filter, with adjacent positive and negative summed
spectra. Those spectra, together with the relevant reference star and arc-lamp spectra, 
were extracted using standard techniques implemented in the {\sl iraf} reduction package. 
All of the stellar spectra were set on a wavelength scale and the individual observations of
each source combined. 

The reference stars are mid-type F dwarfs, generally featureless at near-infrared
wavelengths save for relatively narrow lines in the hydrogen Paschen and Brackett series.
Eliminating those lines from the normalised  spectra 
provides templates which can be used to correct for terrestial absorption features in 
program objects observed at similar altitude and azimuth. The spectra can also be set on
a relative flux scale, since the energy distribution of the high-temperature reference
stars is well represented by a blackbody distribution. There is sufficient overlap between
each of the four passbands employed here to allow accurate cross-calibration, permitting us
to combine the data to form a single spectrum, covering 0.95 to 2.6 $\mu$m.

The flux zeropoint can be determined
using broadband photometry, integrating the spectrum to determine the total flux encompassed
within the 2MASS JHK$_S$ passbands (f$_J$, f$_H$, f$_{Ks}$ erg cm$^{-2}$ sec$^{-1}$). 
Since our spectra are 
on a relative flux scale, this is a two-step process: determining f$_J$, f$_H$ and f$_{Ks}$ 
at magnitude 0; and, given the appropriate conversion factors,  scaling our spectra to match
the broadband photometry listed in Table 1. We have determined the flux/magnitude conversion
factors by applying our analysis to the L2001 observations of 2M0345. This ensures that, as
far as possible,  both sets of observations are tied to the same zeropoint.  
Note that we employ a single scaling factor for each composite spectrum, averaging the 
results from the J, H and K passbands, rather than adjusting each segment separately. In general,
the magnitudes synthesised from the calibrated spectrophotometry agree with the broadband
data to within 3-4\%. The exceptions are the L2 dwarf, 2M1029, 
and the L7.5 dwarf, 2M0825; in both cases, the synthesised H and K magnitudes differ
by +0.1 and -0.1 magnitudes, respectively, from the broadband measurements.
These discrepancies are not important for present purposes.

Figures 1 and 2 plot the final calibrated spectra, where we use our optical spectroscopy to
extend coverage to 6300\AA. All of the latter observations except the data for LP 475-855 
were obtained 
as part of our Keck 2MASS follow-up program using LRIS (Oke {\sl  et al.}, 1995), and the
flux-calibrated spectra span 6300-10100\AA\ at a resolution of 9 \AA.  Full
details of the reduction and analysis are given by Kirkpatrick {\sl  et al.} (1999, 2000). 
 The Keck optical spectra overlap with UKIRT J1 observations, providing an independent 
check on the photometrically-defined flux zeropoints. LP 475-855  was observed
using the modular spectrograph on the Du Pont 100-inch telescope at Las Campanas observatory.
This spectrum covers the wavelength range 6000 to 9500\AA\ at a resolution of 12 \AA. Our observation
happened to catch the star during a flare outburst, which accounts for the strong 
(44\AA\ equivalent width) H$\alpha$ emission. 

Two of the L dwarfs included in our sample, 2M0345 and Denis0205, were 
also observed at near-infrared wavelengths by Leggett {\sl  et al.} (2001). The
L2001 data were also obtained with CGS4, using a nearly identical 
configuration to our own observations. There are, however, differences in the broadband 
photometry adopted for each dwarf, notably in the J-band: UKIRT photometry gives values
of J=13.92 for 2M0345 and J=14.43 for Denis 0205; the corresponding 2MASS data are 
J=14.03 and J=14.55. This at least partly reflects differences in the transmission function of the
respective J filters (note, however, that Dahn {\sl  et al.} (2000) find J=14.59 for Denis 0205
from USNO observations). The 2MASS and L2001 H and K photometry are in  closer agreement. We use
the 2MASS observations and transmission curves to calibrate our data. 

Figure 3 compares the calibrated spectra. There is good agreement
at most wavelengths in the case of the L0 dwarf, 2M0345, with the exception of the 
the 1.0 to 1.1$\mu$m region. The L2001 optical data, from Kirkpatrick, Beichman \& Skrutskie (1997), are
in good agreement with our Keck data. The Denis 0205 observations are more discrepant, 
notably between 0.85 and 1.05 $\mu$m and beyond 2.35$\mu$m. At longer wavelengths, our data are
flatter than the L2001 observations. Our observations of 2M0825
also flatten in the 2.3 to 2.4 $\mu$m range, but the L8 dwarf, 2M0310,
has a steeper energy distribution. 
 
Considering the discrepancy at shorter wavelengths, the L2001 optical data are from Tinney {\sl  et al.}
(1998), with the flux zeropoint set by scaling the observations to match photometry in the
0.8$\mu$m I-band and $\sim1 \mu$m z-band (Leggett, priv. comm.). In contrast, all our optical 
spectroscopy is 
calibrated directly against spectrophotometric standards. The two datasets are in good agreement for
$\lambda < 0.8$ $\mu$m, suggesting that the discrepancy may lie in the L2001 scaling to the z-band 
photometry. Our optical
spectrum of Denis 0205 exhibits no anomalies when compared with our extensive observations of
other L dwarfs, and we have no reason to doubt its reliability. Fortunately, neither of the
long wavelength nor 1$\mu$m  discrepancies is significant for the present analysis. However, 
we use limited wavelength coverage in constructing spectroscopic 
indices to minimise the effect of any residual flux-calibration uncertainties.

\section{ Infrared spectra and spectral classification}

Spectral types serve as a form of astronomical shorthand notation, 
communicating in simple terms the overall appearance of a complex energy 
distribution. The emergent spectrum depends  on the underlying stellar physics, and
the expectation is that  variations in spectral type reflect 
corresponding changes in significant 
physical parameters, notably effective temperature.
A key tenet of any classification scheme must be that it 
rests solely on spectral morphology, independent of 
properties inferred from theoretical models. 
Spectral types are defined by the relative strengths of specific atomic and
molecular features - what the spectrum looks like. Relating those  
observed characteristics to physical parameters such as temperature, chemical abundance
and gravity - what the star {\sl is} -  is a separate, and subsequent, operation.  
A classification system tied directly to physical parameters
must continuously reinvent itself to follow the twists
and turns of the latest theoretical models, and is therefore of limited  value. 

On the other hand, a spectral classification system should also be practical.
The L dwarf spectral sequence, L0 to L8, has been defined using intermediate-resolution
spectroscopy between 6300 to 10100 \AA\ (K99)
\footnote{Mart\'in {\sl  et al.} (1999) have defined an alternative classification scheme,
tied directly to a temperature scale predicted by a specific set of theoretical models.
This {\sl modus operandi}, together with the fact that objects with morphologically 
distinct appearances can be classified as the same spectral type, disqualifies their  
scheme, in our view.}
Few L dwarfs have magnitudes brighter than I=19,  making it difficult to
obtain high signal-to-noise optical observations   except  with the largest available telescopes. 
In contrast, the much higher flux levels in the 1 to 2.5$\mu$m range render these dwarfs
more accessible to spectroscopy at those wavelengths. Indeed, since the bulk of
energy emitted by these cool dwarfs emerges at near-infrared wavelengths, it is important
to ensure that the optically-defined spectral classification scheme currently in place
orders L dwarfs in a manner consistent with the behaviour at 1 to 2.5 $\mu$m. 
 
Previous spectroscopy, notably by Jones {\sl et al.} (1996), Viti \& Jones (1999)
and McLean {\sl  et al.} (2000), shows that late M and
L dwarfs possess an array of near-infrared atomic lines and molecular bands,  suitable
for spectral classification. The strongest of those features are discussed further in
the following section,  where we show that
the behavioural trends are compatible with variation in the optical spectral type. 
Given this consistency, there is no requirement to define an independent near-infrared
spectral classification system. Combining the present sample with
the late-type dwarfs observed by  L2001 gives a total of eighteen dwarfs with spectral
types of M8 or later, all with optically-determined spectral types; those ultracool
dwarfs provide an excellent means of defining an infrared classification scheme which is
tied to the established K99 L dwarf spectral sequence. 

\subsection {Photometric properties}

Table 2 lists the available IJHK photometry for the sample of ultracool dwarfs
considered here. Where possible, we list JHK observations from the USNO program (Dahn 
{\sl  et al.}, 2000), since those data are  closer to the UKIRT system
used by L2001. As noted above, there can be differences of up to 0.1
magnitude with respect to the 2MASS data listed in Table 1, particularly in the J band. 
Twelve sources have accurate trigonometric parallax measurements,
primarily from Dahn {\sl  et al.} (2000), and those data are also listed. 

Figures 4 and 5 place these measurements in the broader context, plotting 
the (M$_J$, (I-J)) colour-magnitude diagram, and the far-red/near-infrared colour-colour 
and spectral-type/colour relations outlined by nearby stars and brown dwarfs. 
Our targets lie within the late-M/L-dwarf distribution in most of these
diagrams; the exception is the (spectral type, (J-K$_s$)) relation, where the
L3 to L7 dwarfs are among the bluer objects in their
respective classes. This holds regardless of whether we use exclusively 2MASS photometry or the
data listed in Table 2. 

We have integrated the spectral energy distributions to determine bolometric magnitudes
for the dwarfs in our sample. Following L2001, we allow for short wavelength
radiation by linearly extrapolating from our shortest wavelength datum to F$_\lambda = 0$
at $\lambda = 0$; at long wavelengths, we assume that the flux distribution follows a
Rayleigh-Jeans distribution. Between 65 and 75\% of the total energy radiated by these
ultracool dwarfs emerges between 1 and 2.5$\mu$m; as L2001 emphasise, 
longer wavelengths account for most of the remainder. 

Table 2 lists the  bolometric corrections calculated for the J-band,
where we also give the values derived from L2001. 
These results are plotted against spectral
type in the lowest panel of Figure 5. As previously suggested (Reid {\sl  et al.}, 2000), 
there is relatively little variation in the J-band bolometric corrections of
dwarfs cooler than Wolf 359 (spectral type M6); 
indeed, most of the dwarfs considered here have values of BC$_J$ within
0.5 magnitudes of Leggett {\sl  et al.'s} (1999) result for Gl 229B. This relatively small
spread in bolometric correction is in contrast to behaviour in
the  H and K passbands, where the onset
of methane absorption leads to changes exceeding 1.5 magnitudes. 
M$_J$ can therefore provide a good guide to total luminosity even in the absence of
detailed spectroscopic information.
The implications of these results for the likely temperature scale of the L dwarf sequence
will be discussed in conjunction with near-infrared observations of a larger sample of
L dwarfs by Gizis {\sl  et al.} (in prep.). 

\subsection {Near-infrared atomic and molecular features}

Near-infrared spectra of ultracool dwarfs are dominated by the broad
absorption bands at 1.4 and 1.85$\mu$m due to H$_2$O. Detailed spectroscopy, however,
reveals numerous other features.  Early and mid-type M dwarfs exhibit numerous
atomic lines, principally in the J and K windows, due to the alkaline elements (Na, Ca, K),
iron, titanium and silicon (Mould, 1978; Kleinmann \& Hall, 1986; Davidge \&
Boeshaar (1991)). Observations of the M8 dwarfs TVLM 513-46546
and Gl 569B by Viti \& Jones (1999) show that
while the atomic lines remain prominent in the J passband, 
the longer wavelength features weaken at these later spectral types. 

Figures 6, 7 and 8 present  J, H and K-band spectra of the eighteen ultracool
dwarfs considered here. We have identified the more prominent atomic features,
including the two K I doublets, at 1.169/1.177 and 1.244/1.252 $\mu$m
and the Na I doublet at 1.138/1.140 $\mu$m. The last mentioned is unresolved at the
resolution of the current spectra, 
and can be distorted by terrestial H$_2$O absorption.
An absorption line due to the blended Na I 2.206/2.209 $\mu$m doublet 
is detected in  some spectra, while narrow lines, possible due to Al I, 
are also evident at $\sim1.3 \mu$m in LHS 2065 and GD 165B. 
Table 3 lists measurements of the equivalent widths of
these features, including the strongest FeH bandhead
(possibly blended with VO) at 1.2$\mu$m. The wings of the KI 1.244/1.252 $\mu$m
doublet overlap at the resolution of our observations, so we give the joint equivalent
width. The uncertainties in the measurements vary, dependent on the signal-to-noise of
the individual spectra, but are typically 1 to 2 \AA.

None of the CGS4 spectra show evidence for the 1.71$\mu$m Mg I feature 
detected in TVLM 513-46546 by Viti \& Jones (1999), However, 
inspection of the spectral sequence in figure 7 suggests 
the presence of recurrent features near the peak of the H-band flux distribution.
Those features are present in both our own observations and the L2001 dataset, and
may also be evident in Mould's (1978) observations of Gl 699 and in TVLM 513-46546
(Figure 2 in Viti \& Jones, 1999). Figure 9 shows an
expanded plot of the region, including four L dwarfs with relatively high signal-to-noise
data, and the normalised spectrum of the F6 dwarf, BS1358. Three possible features in
the L dwarf spectra, at 1.58, 1.613 and 1.627 $\mu$m, are identified; at least the first
mentioned also appears to be present in M8 dwarfs. All three are
close to, but not exactly coincident with, terrestial absorption  in the 
BS1358 spectrum. However, the features appear to increase in strength from M8 to L4/L5,
declining rapidly to the threshold of detection in the L7 dwarf Denis 0205. This
is consistent with an intrinsic origin, and the
absorption pattern is suggestive of molecular absorption.
Higher resolution, high signal-to-noise spectra at these wavelengths are required to
confirm the nature of these features. 

The central regions of the near-infrared steam bands are masked by terrestial 
absorption, even in observations from Mauna Kea. The higher
temperatures in late-M and L dwarf atmospheres, however, lead to broader bands,
placing the wings in regions accessible to observation. H$_2$O absorption 
increases in strength through spectral type M, and Mart\'in {\sl  et al.} (1999)
find that trend to continue through mid-type L dwarfs, at least at 1.4$\mu$m.
We have constructed  four indices designed to measure the depth of the
1.4 and 1.85 $\mu$ bands. 
The indices are defined as
\begin{displaymath}
H_2O^A \ = \ {F_{1.34} \over F_{1.29}} \qquad \qquad H_2O^B \ = \ {F_{1.48} \over F_{1.60}}
\end{displaymath}
\begin{displaymath}
H_2O^C \ = \ {F_{1.70} \over F_{1.80}} \qquad \qquad H_2O^D \ = \ {F_{2.0} \over F_{2.16}}
\end{displaymath}
In each case, the flux measurement is defined as the average flux in a 0.02 $\mu$m window
centred on the wavelength in question; that is, $F_{1.34}$ is the average flux in interval
$1.33 < \lambda < 1.35 \ \mu$m.
Note that $F_{1.34}$, in particular, lies in close proximity to a region affected by
 terrestial H$_2$O absorption.

The triple bandheads produced by the first overtone transitions of the CO molecule 
are also evident at 2.3 $\mu$m in the K passband, with the longer wavelength transitions
becoming less distinct with increasing spectral type and barely detectable beyond L5.
We have measured a bandstrength for this feature by constructing the R$_{CO}$ index, 
defined as the ratio between the flux at 
the base of the primary bandhead (at 2.29 $\mu$m) and the pseudo-continuum flux at 2.27$\mu$m.
Tokunaga \& Kobayashi (1999) have shown that H$_2$ absorption is also present 
at $\sim2.2$ $\mu$m in later-type L dwarfs. They define two indices,
\begin{displaymath}
K1 \ = \ {{\langle F_{2.10-2.18}\rangle - \langle F_{1.96-2.04}\rangle} \over
{0.5(\langle F_{2.10-2.18}\rangle + \langle F_{1.96-2.04}\rangle)}}
\end{displaymath}
and
\begin{displaymath}
K2 \ = \ {{\langle F_{2.20-2.28}\rangle - \langle F_{2.10-2.18}\rangle} \over
{0.5(\langle F_{2.20-2.28}\rangle + \langle F_{2.10-2.18}\rangle)}}
\end{displaymath}
K1, like H$_2$O$^D$, provides a measure of the red wing of the 1.85$\mu$m steam band, while K2 
is sensitive to the presence of H$_2$ absorption. 

None of the spectra, including our observations of the L8 dwarf, 
2M0310, show evidence for methane absorption in either the H or K band.
However, the J-band spectrum of 2M0825 clearly shows the onset of broad
H$_2$O absorption at 1.1 to 1.2 $\mu$m, a prominent feature in the
early-type T dwarfs discovered by Leggett {\sl  et al.} (2000). 
Table 4 lists the molecular indices for the eighteen ultracool dwarfs
in the current sample. The measured ratios are typically uncertain to $\pm$5\%.

\subsection {Spectral type calibration at near-infrared wavelengths}

Figure 10 plots the variation with spectral type of the equivalent widths of the atomic
sodium and potassium lines present in the J-band. In both this figure and the succeeding figures, we
represent spectral type numerically, setting -2$\equiv$M8, 0$\equiv$L0, 5$\equiv$L5 and so forth.
All of the equivalent widths show significant
dispersion at a given spectral type, at least partially due to the measurement uncertainties. 
The potassium lines reach maximum strength at type L3/L4, and decline rapidly at later types.
The  sodium blend increases in strength from M8 to L5, but is ill defined at
later spectral types. We note also that the 2.20 $\mu$m Na I line is not detected in
most mid-type L dwarfs, but appears to be present in 2M0036 and Denis 0205. It is not clear
whether this is an intrinsic effect or more a reflection of signal-to-noise in the various
spectra.

Of the molecular indices, the R$_{CO}$ ratio is essentially unchanged over the full 
spectral range. In similar fashion, most dwarfs show little evidence for H$_2$ 
absorption. The upper panel of Figure 11 plots Tokunaga \& Kobayashi's (1999)
K1 and K2 indices; 
as pointed out by those authors, Denis 0205 stands out from the main body, with
a significantly-depressed K2 index. 2M0310 lies even further below the mean, 
although the L7.5 dwarf, 2M0825, is unremarkable in this diagram. Inspection of the K-band
spectra plotted in Figure 8 suggests substantial H$_2$ absorption is present in 2M0310.

The lower panel in Figure 11 plots the K1 index against optical spectral type. The data
are well correlated for spectral types earlier than L6, and least-squares fitting
gives the relation
\begin{displaymath}
Sp \ = \ (-2.8 \pm 0.6) \ + \ (21.8 \pm 2.8) K1,  \qquad \sigma \ = \ \pm 0.95
\end{displaymath} 

Figure 12 plots the four H$_2$O  indices defined in the previous section.
All are well correlated with optical spectral type, although both
H$_2$O$^A$ and H$_2$O$^D$ ``saturate'' at later spectral types. 
H$_2$O$^A$ and  H$_2$O$^B$ show the least dispersion, with
\begin{displaymath}
Sp \ = \ 23.4\pm2.7 \ - \ (32.1\pm4.2) H_2O^A \ , \qquad \sigma \ = \ \pm1.18
\end{displaymath}
where the fit is valid for spectral types earlier than L6, and 
\begin{displaymath}
Sp \ = \ 20.7\pm1.6 \ - \ (24.9\pm2.2) H_2O^B \ , \qquad \sigma \ = \ \pm1.02
\end{displaymath}
for spectral types M8 to L8. 
Combined with the K1 calibration given above, these relations bridge the 
optical/infrared divide;  infrared spectroscopy can be used to determine 
spectral types consistent with optically-defined K99 system.

\section {Summary and conclusions}

We have presented near-infrared spectra for a sample of eleven ultracool dwarfs. 
Our observations complement the dataset compiled
previously by Leggett {\sl  et al.} (2001), providing good coverage of optically-defined spectral 
types from M8 to L8. We have combined these near-infrared data with optical spectra,
integrating the energy distribution to derive bolometric magnitudes. Our results
confirm previous suggestions that there is relatively little variation in the
J-band bolometric correction for cool dwarfs ranging from spectral type $\approx$M6 to
type T (Gl 229B). 

Combining the two datasets, we have measured equivalent widths for the stronger atomic
lines due to potassium and sodium. The results show that the variation of spectral
features between 1 and 2.5 $\mu$m  is consistent with the behaviour at far-red wavelengths;
specifically, we can map the infrared variation onto the Kirkpatrick {\sl  et al.}
(1999) spectral classification scheme. The majority of the atomic lines show similar
characteristics, increasing in equivalent width to maximum at L3/L4, with a more
rapid decline in strength at longer wavelengths. The exception is the 2.2$\mu$m Na I line,
although its intermittent detection may simply reflect the signal-to-noise of the observations.
Inspection of the  combined suite of near-infrared spectra suggests that there are
intrinsic absorption features between 1.58 and 1.63 $\mu$m which remain to be identified. 

Besides measuring equivalent widths for atomic lines, we have constructed narrowband
indices designed to measure the strength of the prominent molecular absorption bands
due to water and CO. Even the latest-type dwarf observed, spectral type L8, shows no
evidence for methane absorption. There is little variation in CO absorption, but all
four H$_2$O indices exhibit a good correlation with spectral type, with the
best calibration provided by H$_2$O$^B$, measuring the depth of absorption in the
redward wing of the 1.4$\mu$m steam band. 

Our analysis is based on the most extensive set of near-infrared spectra currently available.
Even so, we have data for only fourteen L dwarfs, four of which have incomplete spectral
coverage. Observations of a larger sample of ultracool dwarfs are required to verify these results.
However, the correlations derived here not only permit 
extension of the K99 spectral sequence to near-infrared wavelengths, 
but also indicate that spectral type on that system is monotonically correlated with changes in
the effective temperature of these low-mass dwarfs.

\acknowledgments { }
We would like to thank Thor Wald and Tom Kerr  for assistance in obtaining
our observations and with the initial data reduction.  We also thank Sandy Leggett for making
the L2001 spectra readily available in electronic form.  INR and JDK  acknowledge
partial support from a NASA/JPL grant to 2MASS Core Project Science. JEG and JDK 
acknowledge the support of the Jet Propulsion Laboratory, California Institute of
Technology, which is operated under contract wth NASA.
The United Kingdom Infrared Telescope is
operated by the Joint Astronomy Centre on behalf of the Particle Physics and Astronomy
Research Council.

\newpage

\begin{table}
\begin{center}
{\bf Table 1: Observing Log}
\begin{tabular}{lcccrrrrr}
\tableline\tableline
2MASS & Sp. & J &  J-K$_S$  & J1 & J2 & H & K & reference \\
\tableline
IJ0429028+133759 & M7 & 12.67  & 1.03  & 960s & 960s & 960s & 960s & HR 1358, F6 \\
WJ0320596+185423 & M8 & 11.74  & 1.17  & 960  & 960 & 640 &  960 & HR 869, F6 \\
PJ0345432+254023 & L0 & 14.03  & 1.33  & 2880    & 2880   & 2880   &  1920 & HR 1238, F4 \\
WJ0746425+200032 & L0.5 & 11.74  & 1.25  &960& 960 & 640 & 1280 & HR 2835, F6 \\ 
WJ0829066+145622 & L2 & 14.72  & 1.60  & \null & 2400 & 1920 & 1280 & HR 3299, F6 \\
IJ1029216+162652 & L2.5 & 14.31  & 1.70  & \null & 1440 & 960 & 960 & HR 3998, F7 \\
WJ0036159+182110 & L3.5 & 12.44  & 1.41  & 960& 960 & 960 & 1280 & HR 217, F8 \\ 
IJ1112256+354813 &L4.5 &  14.57  & 1.77  & \null  & 1600 &1600   & 1280 & HR 4412, F7 \\  
DENIS-P J0205.4-1159 & L7 & 14.55  & 1.56  & 3840 & 3840 & 2240 & 1280 & HR 638, F5 \\ 
WJ0825196+211552 & L7.5 & 15.12  & 2.07  & 3840 & 3840  &2240 & 2560 & HR 3299, F6 \\
WJ0310599+164816 & L8 & 16.43  & 2.03  &\null   & 4320  & 2240  &   1920 & HR 869, F6 \\
\tableline\tableline
\end{tabular}
\end{center}
Notes: Columns 3 and 4 list near-infrared photometry from 2MASS. \\
Columns 5 to 8 list the total integration time (in seconds) for each of the 
four wavelength ranges described in the text. \\
Column 9 identifies the reference star used to calibrate each target. \\
    2MASSIJ0429028+133759 is  LP 475-855 (Leggett \& Harris, 1994)\\
    2MASSWJ0320596+185423 is the proper motion star LP 412-31 \\
    2MASSIJ1112256+354813 is also known as Gl 417B 
\end{table}

\clearpage

\begin{table}
\begin{center}
{\bf Table 2:  Photometric properties}
\begin{tabular}{lccccccccc}
\tableline\tableline
Name & Sp. type & J & (I-J)  & (J-H) & (H-K$_s$) & BC$_J$ & $\pi$ (mas)  & source\\
\tableline
2M0320/LP412-31 & M8 & 11.74 & 3.05 & 0.71 & 0.51&1.97  & 68.0$\pm$0.7 & USNO \\
TVLM 513-46546 & M8.5 & 11.80 & 3.41 & 0.68 & 0.40&2.05  & 94.3$\pm$0.7 & USNO\\
LHS 2065 & M9 & 11.22& 3.22 & 0.80 & 0.48 &1.90 & 117$\pm$2 & L2001\\
BRI0021-0214 & M9.5 &  11.80 & 3.39 & 0.76 & 0.46 & 1.95 &$82\pm2^C$ & L2001\\
2M0345 & L0 & 13.92 &3.44 & 0.76 & 0.54 &2.01 & $36.7\pm0.7$ & USNO\\
2M0746$^A$ & L0.5 &  11.73 & 3.38 & 0.77 & 0.47 &2.01& $82.6\pm2$ & USNO\\
2M0829 & L2 & 14.72 & \null & 0.93 & 0.67 &1.66 & \null & 2MASS\\
2M1029 & L2 & 14.31 & \null & 0.96 & 0.74 &1.92 & \null & 2MASS\\
Kelu 1$^B$ & L2 & 13.42 & 3.52 & 0.95 & 0.67 &1.67 & $52.1\pm2.0$&  USNO \\
Denis 1058 & L3 & 14.16 & 3.64 & 0.88 & 0.63 & 1.76 &$57.0\pm1.2$ & USNO \\
2M0036 & L3.5 &  12.44 & 3.67 & 0.84 & 0.54 &1.99& $112.4\pm2.0$ & USNO\\
GD 165B & L4 &  15.71 & 3.45 & 1.03 & 0.55 &1.69 & $31\pm2$ & L2001\\
2M1112$^B$ & L4.5 & 14.57 & \null & 1.10 & 0.78 &1.61 & 46.0$\pm0.8^D$ & 2MASS \\
Denis 1228$^{A,B}$ & L5 &  14.37 & 3.84 & 0.97 & 0.66& 1.66 &$51\pm3$ & USNO\\
SDSS 0539 & L5 &  13.94 & 3.73 & 0.97 & 0.53 & 1.76 &\null & L2001\\
Denis 0205$^A$ & L7 & 14.59 & 3.85 & 0.97 & 0.56 & 1.71 &$55.5\pm2.3$ & USNO \\
2M0825$^B$ & L7.5 & 14.99 & 4.23 & 1.17 & 0.80 & 1.70 &\null & USNO\\
2M0310$^B$ & L8 &  16.43 & \null & 2.03 & 1.48 & 1.66& \null & 2MASS\\
\tableline
\tableline\tableline
\end{tabular}
\end{center}
Source: USNO - IJHK Dahn {\sl  et al.} (2000) \\
2MASS - I, Dahn {\sl  et al.} (2000); JHK$_S$, 2MASS \\
L2001 - IJHK, Leggett {\sl  et al.} (2001) \\
Notes: \\
A. known L-dwarf/L-dwarf binary \\
B. Li 6708\AA\ detection implies $M \le 0.06 M_\odot$ \\
C. Parallax from Tinney (1996) \\
D. Parallax from Hipparcos astrometry of Gl 417A
\end{table}

\clearpage
\begin{table}
\begin{center}
{\bf Table 3:  Equivalent width measurements}
\begin{tabular}{lcccccccc}
\tableline\tableline
Name & Sp. type & Na I & Na I  & K I &K I &K I & FeH  \\
     &    & 1.14$\mu$m & 2.2$\mu$m & 1.169$\mu$m & 1.197$\mu$m & 1.24/1.25$\mu$m & 1.2$\mu$m \\
\tableline
LP 412-31 & M8 & 20.2 &  10.1 & 5.7 & 6.6 & 17.8& 16.3  \\
TVLM 513-46546 & M8.5 & 21.5 &3.3 &  6.4 & 8.6 & 22.6 &  19.0 \\
LHS 2065 & M9 & 18.7 & 3.0 & 6.2 &  7.5 & 19.0 & 18.3 \\
BRI0021-0214 & M9.5 & 20.8 & 5.7 & 5.4 & 7.1 & 21.8 & 22.6 \\
2M0345 & L0 & 14.8 & \null & 9.5 & 8.8 & 21.0 & 18.8 \\
2M0746 & L0.5 & 19.4 & \null & 7.8 & 10.5 & 23.6 &  19.2 \\
2M0829 & L2 & 16.1 & \null & 6.8 & 9.3 & 25.6 & 23.0 \\
2M1029 & L2 & \null & \null &  \null & \null & 24.2 & \null \\
Kelu 1 & L2 & 22.4 & \null & 8.8 & 8.9 & 25.3 & 20.1 \\
Denis 1058 & L3 & 23.5 & \null & 10.8 & 16 & 46.8 & 27.2 \\
2M0036 & L3.5 & 27.3 & 3.4 & 10.9 & 14.0 & 37.4 & 23.1 \\
GD 165B & L4 & 32.2 & \null & 8.3 & 9.8 & 28.6 & 19.0 \\
2M1112 & L4.5 & \null & \null &  \null & \null & 29.9 & \null \\
Denis 1228 & L5 & 20.7 & \null & 7.1 & 10.1 & 20.9 & 16.7 \\
SDSS 0539 & L5 & 23.9 & \null & 9.2 & 11.3 & 22.9 & 18.6 \\
Denis 0205 & L7 & 24.8 & 3.9 & 8.6 & 8.3 & 16.8 & \null \\
2M0825 & L7.5 & \null & 8.2 & 4.8 & 4.2 & 8.5 \\
2M0310 & L8 & \null & 10: &  \null & \null & 9.0 & \null \\
\tableline
\tableline\tableline
\end{tabular}
\end{center}
Equivalent widths given in \AA ngstroms. 
\end{table}

\clearpage

\begin{table}
\begin{center}
{\bf Table 4: Molecular band indices}
\begin{tabular}{lcccccccc}
\tableline\tableline
Name & Sp. type & H$_2$O$^A$  & H$_2$O$^B$ & H$_2$O$^C$ & H$_2$O$^D$ & K1 & K2  & R$_{CO}$ \\
\tableline
LP 412-31 & M8 &   0.79  &  0.91  &  0.71 &   0.96  &  0.03 &  -0.09 & 0.77 \\
TVLM 513-46546 & M8.5 &  0.71 &  0.87  &  0.64 &   0.89 &   0.10 &  -0.03 & 0.82 \\
LHS 2065 & M9 &    0.77 &   0.88 &   0.71 &   0.88 &   0.11 &  -0.01& 0.82 \\
BRI0021-0214 & M9.5 &   0.70 &   0.82 &   0.66 &   0.82 &   0.16 &  -0.01 &  0.80 \\
2M0345 & L0 &   0.72 &   0.81 &   0.69  &  0.94  &  0.08 &  -0.03& 0.75 \\
2M0746 & L0.5 &    0.68 &   0.79  &  0.63 &   0.87  &  0.12&   -0.08&  0.77 \\
2M0829 & L2 &    0.72  &  0.79  &  0.73  &  0.72  &  0.28  &  0.03 &   0.74 \\
2M1029 & L2 &  0.64 &   0.71 &   0.70 &   0.74 &   0.26 &  -0.03  &     0.72 \\
Kelu 1 & L2 &   0.65  &  0.71  &  0.61  &  0.74 &   0.26  & -0.01  &    0.78 \\
Denis 1058 & L3 &   0.63 &   0.67 &   0.65 &   0.78 &   0.17 &  -0.01 &   0.81 \\
2M0036 & L3.5 &  0.57  &  0.66 &   0.54 &   0.74 &   0.26 &  -0.08 & 0.77 \\
GD 165B & L4 &   0.61  &  0.72  &  0.55  &  0.73 &   0.30 &  -0.04 &     0.76 \\
2M1112 & L4.5 &    0.64  &  0.62 &   0.59  &  0.69  &  0.32 &  -0.10  &   0.75 \\
Denis 1228 & L5 &    0.63 &   0.67 &   0.65 &   0.78  &  0.32 &  -0.08 &   0.77 \\
SDSS 0539 & L5 &  0.54 &   0.65 &   0.51 &   0.71  &  0.32 &  -0.09  &   0.80 \\
Denis 0205 & L7 &  0.56  &  0.57  &  0.46  &  0.70  &  0.34 &  -0.16 &   0.79 \\
2M0825 & L7.5 &   0.62  &  0.64  &  0.63  &  0.69  &  0.33 &  -0.08  &   0.75 \\
2M0310 & L8 &   0.51 &   0.46 &   0.48 &   0.77 &   0.26 &  -0.26    &   0.76 \\
\tableline
\tableline\tableline
\end{tabular}
\end{center}
\end{table}

\clearpage

\begin{figure}
\plotone{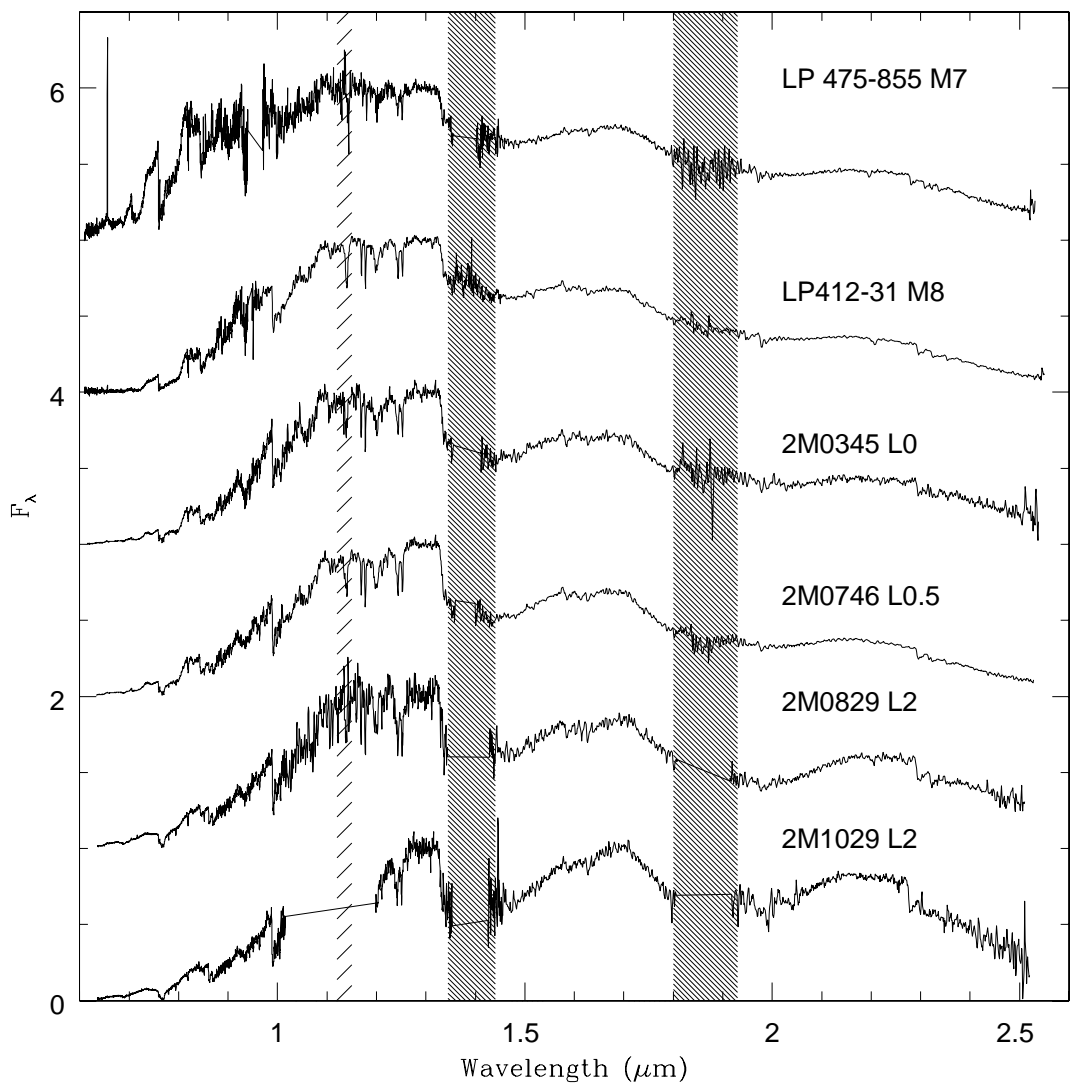}
\caption {Optical and near-infrared spectra of the earlier-type dwarfs in the present sample. 
All of the spectra are normalised to the mean flux at $1.28\pm0.01$ $\mu$m and the
shaded regions denote wavelengths affected by terrestial water vapour absorption; the
lighter-shaded 1.14$\mu$m band is significantly weaker than the 1.4 and 1.85$\mu$m absorption.
The optical data are from our Keck LRIS observations with the exception of LP 475-855, which was
observed using the Modular Spectrograph on the Du Pont 100-inch at Las Campanas observatory.
The last-mentioned star was caught during a flare outburst, accounting for the strong 
emission at H$\alpha$. }
\end{figure}

\clearpage

\begin{figure}
\plotone{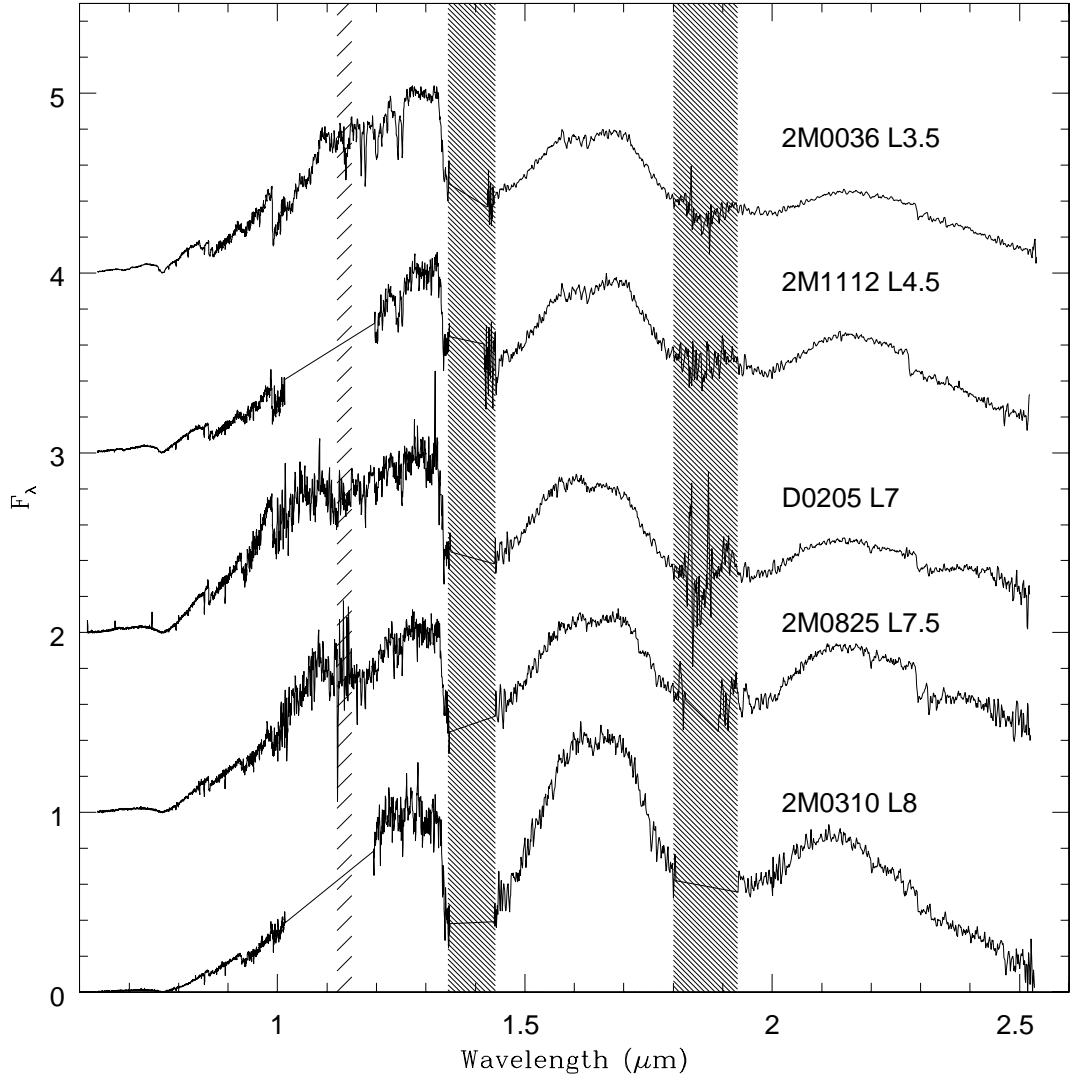}
\caption {Optical and near-infrared spectra of the later-type dwarfs in the present sample.}
\end{figure}

\clearpage

\begin{figure}
\plotone{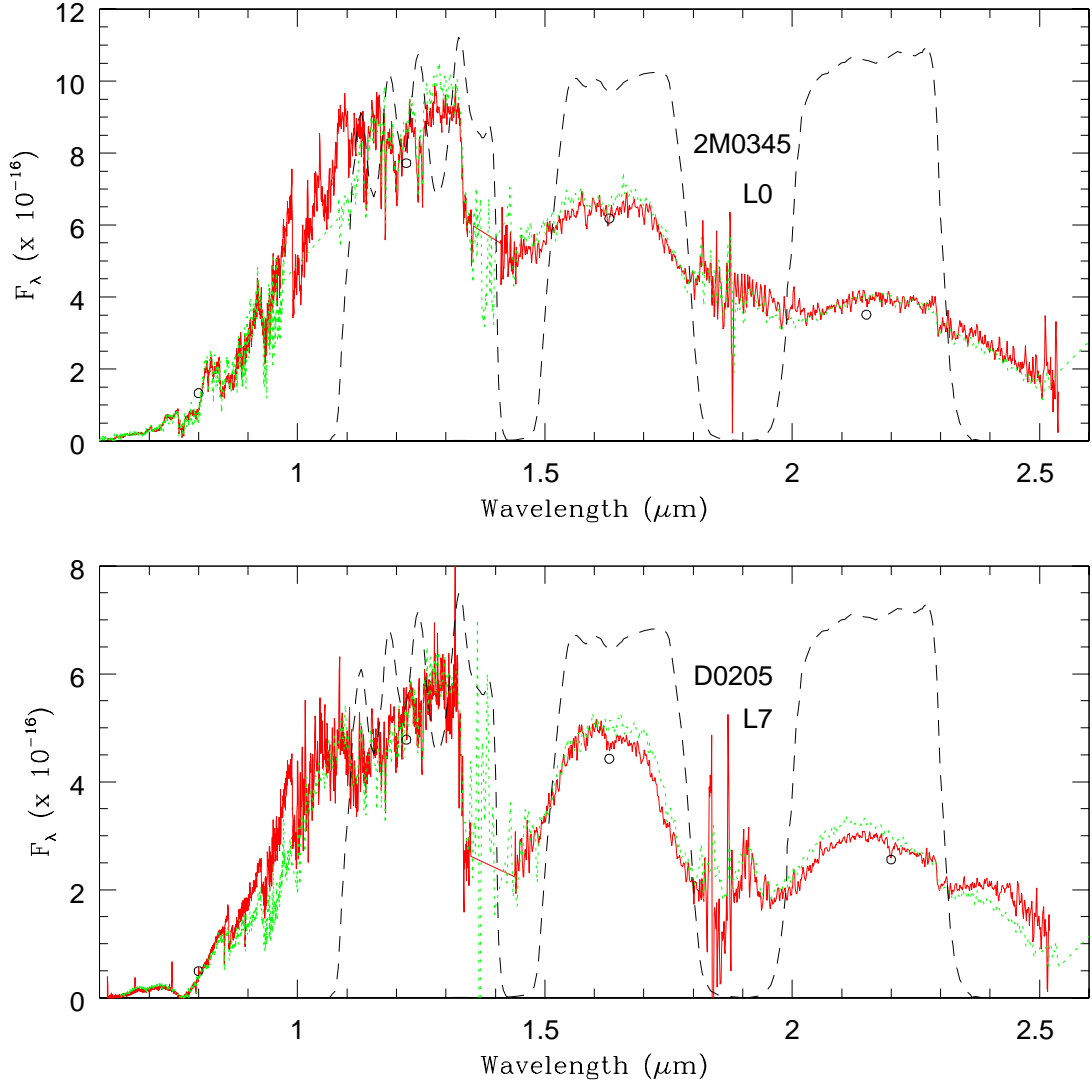}
\caption {Comparison between our observations of 2M0345 and Denis 0205 and the near-infrared
spectroscopy of Leggett {\sl  et al.} (2001). Our data are plotted as a solid line. The 2MASS
JHK$_s$ photometric passbands are superimposed on the data.}
\end{figure}

\clearpage

\begin{figure}
\plotone{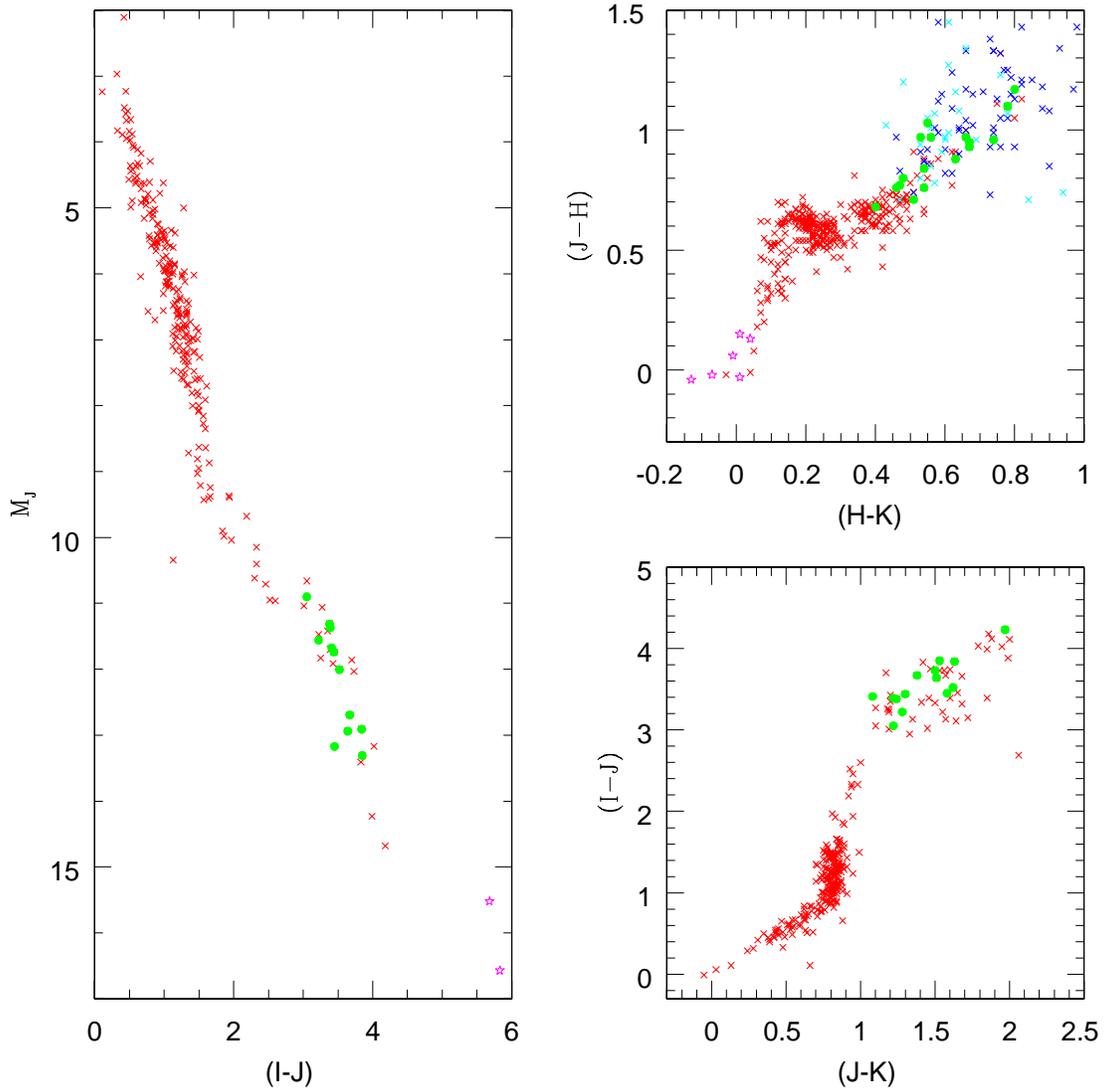}
\caption { Colour-magnitude and colour-colour data for late-type dwarfs. The solid points
identify the ultracool dwarfs with CGS4 near-infrared data - the sources listed in Table 2.
Crosses are nearby stars (photometry from Leggett, 1992)  and L dwarfs, from Kirkpatrick 
{\sl  et al.} (1999, 2000); five-point stars are T dwarfs. }
\end{figure}

\clearpage

\begin{figure}
\plotone{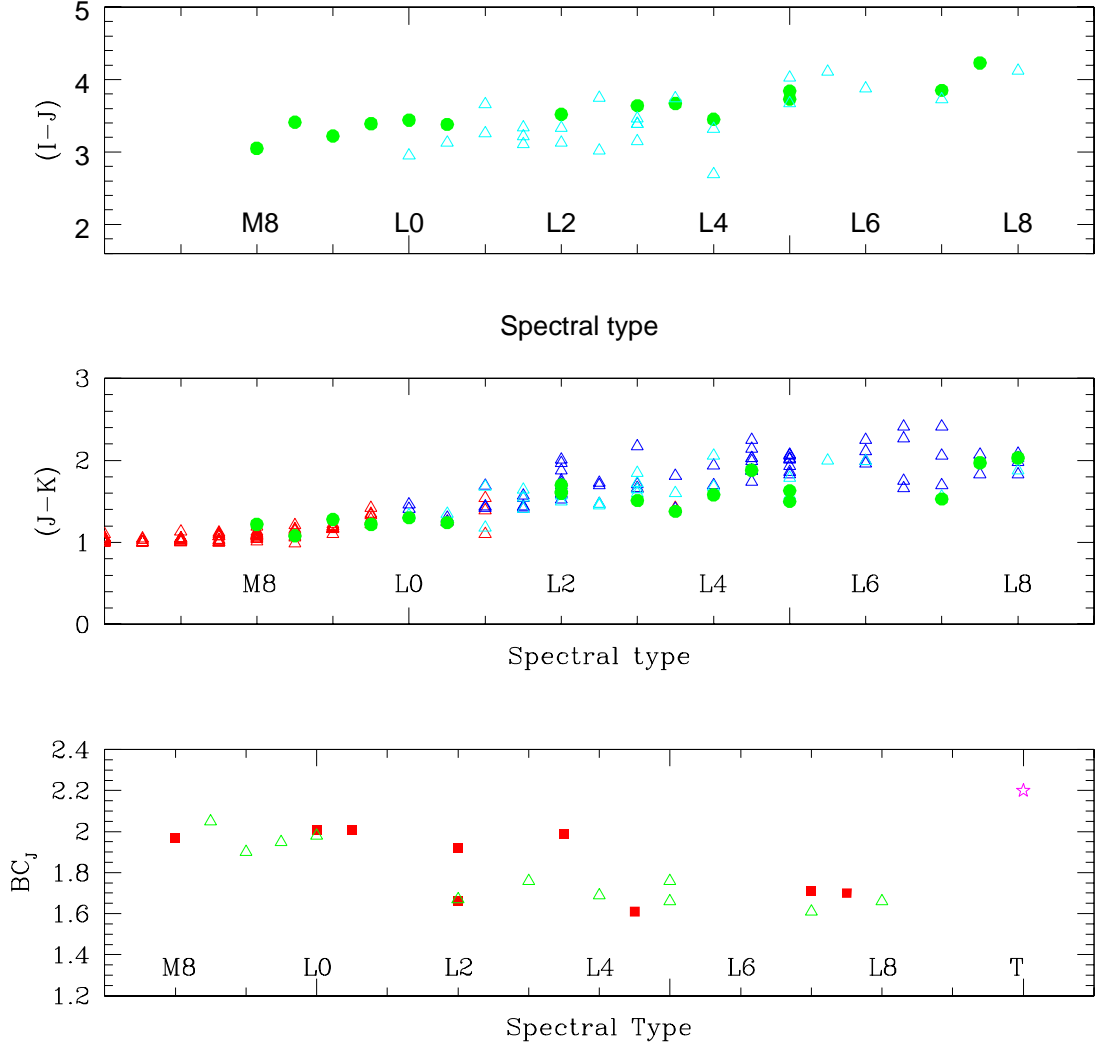}
\caption { Colour/spectral-type relations for ultracool dwarfs. In the upper two panels, open 
triangles are L dwarfs from Kirkpatrick {\sl  et al.} (1999, 2000); solid points plot data
for the dwarfs listed in Table 2. The lowest panel plots the J-band bolometric correction
as a function of spectal type, where the solid squares are from our analysis, while the open
triangles are taken from the L2001 calibration. The five-point star marks Gl 229B.} 
\end{figure}

\clearpage

\begin{figure}
\plotone{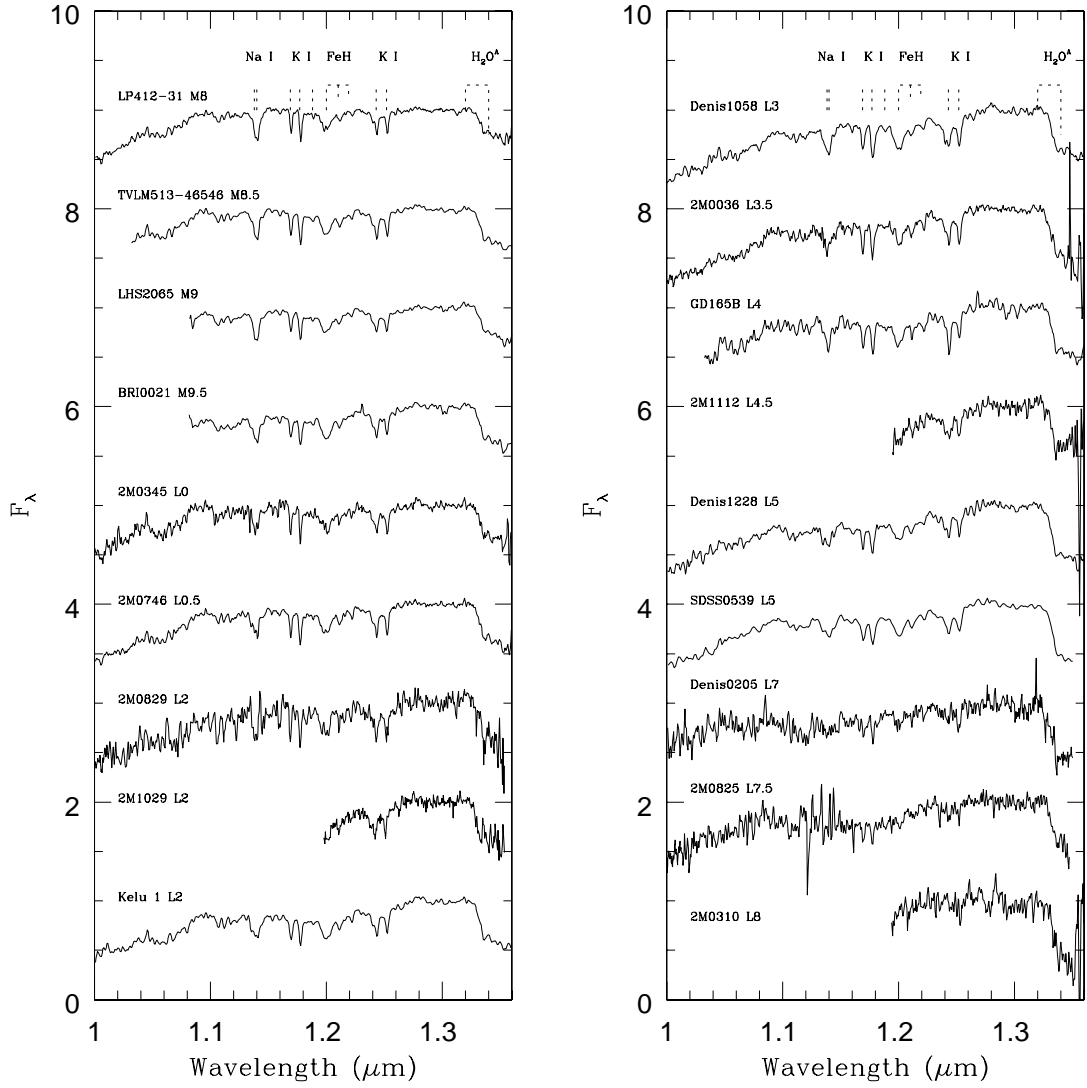}
\caption {J-band spectra of the 18 ultracool dwarfs from L2001 and our own observations.
The more prominent atomic and molecular features are identified.}
\end{figure}

\begin{figure}
\plotone{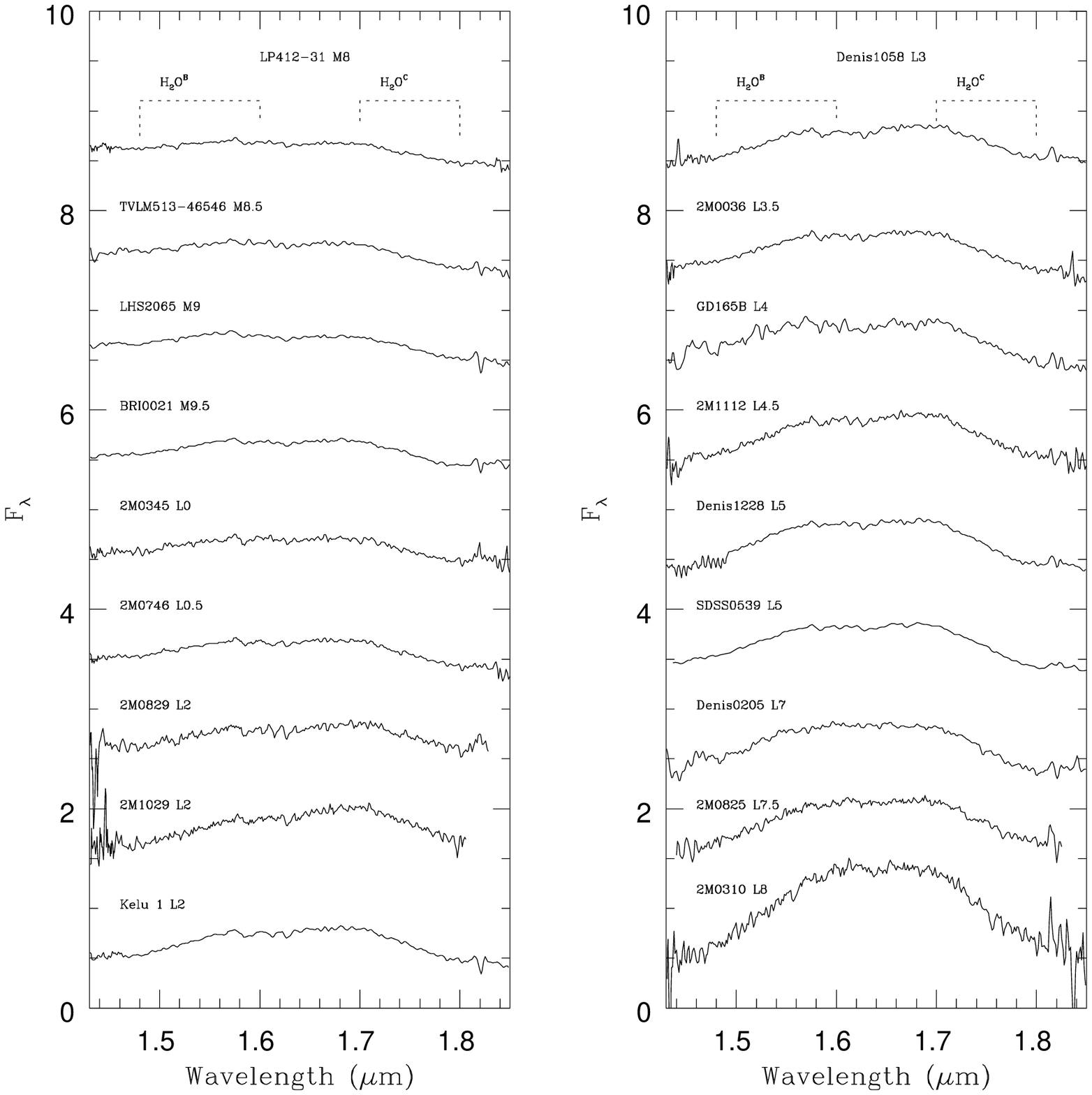}
\caption {H-band spectra of the 18 ultracool dwarfs from L2001 and our own observations.}
\end{figure}

\begin{figure}
\plotone{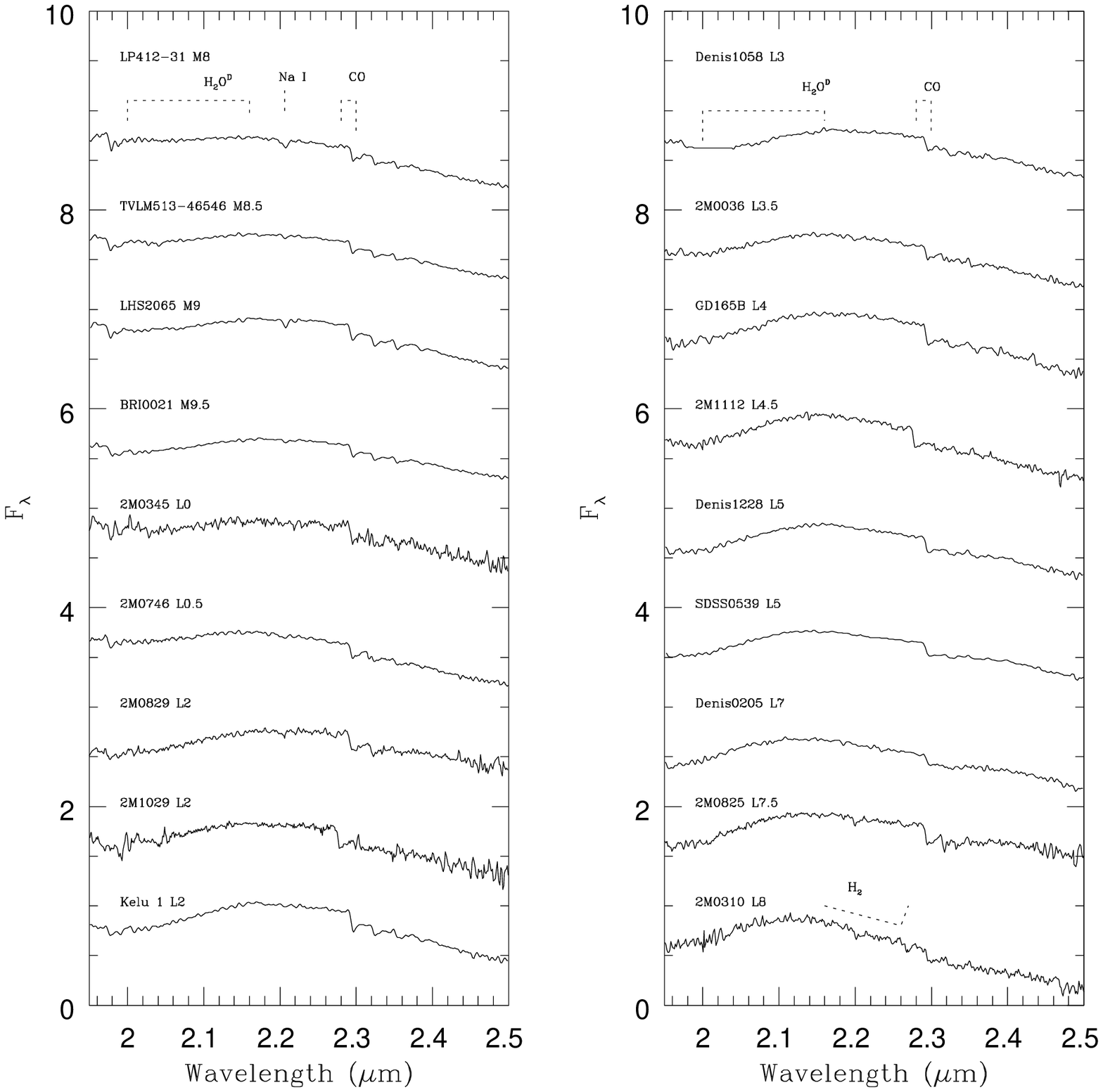}
\caption {K-band spectra of the 18 ultracool dwarfs from L2001 and our own observations.}
\end{figure}

\begin{figure}
\plotone{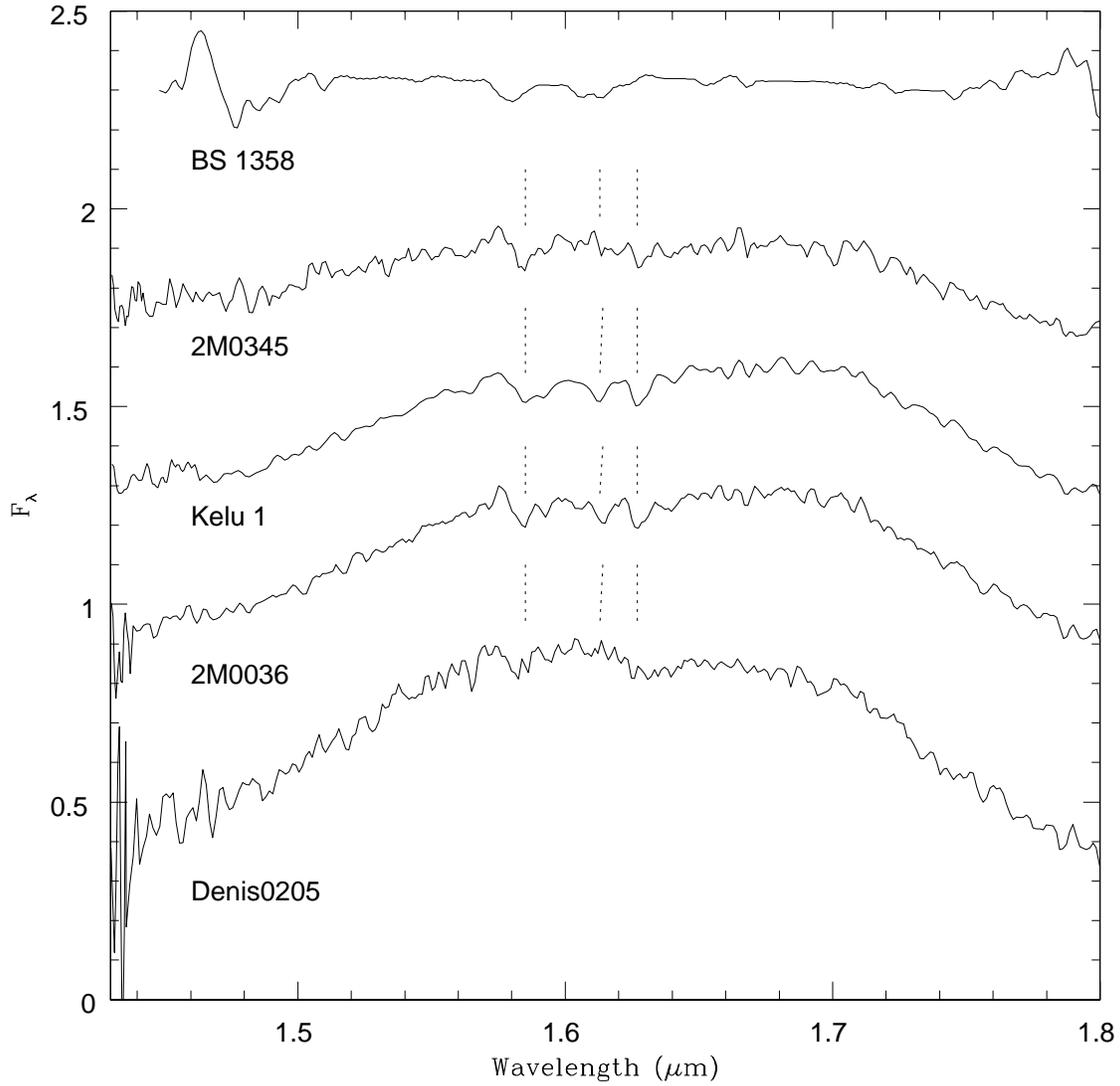}
\caption {An expanded plot of the H-band region in four L dwarfs, 2M0345 (L0), Kelu 1 (L2),
2M0036 (L3.5) and Denis0205 (L7). The possible atomic or molecular features discussed in the text
are marked, and the normalised spectrum of BS 1358 plotted for comparison. The 
``feature'' at 1.4$\mu$m in the standard star spectrum is a product of the continuum fitting, and
does not affect calibration of the program objects.} 
\end{figure}

\begin{figure}
\plotone{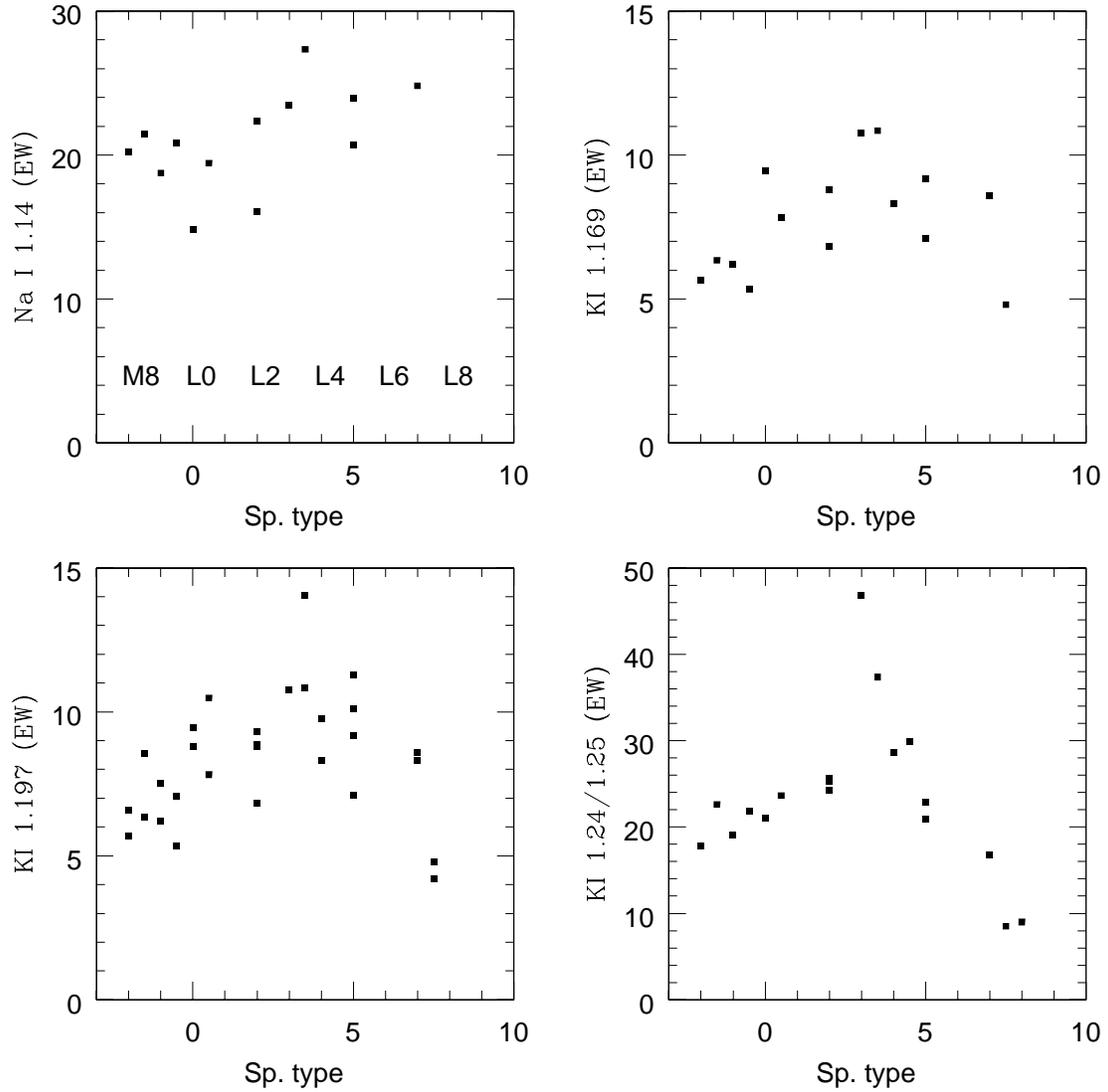}
\caption{ The variation in equivalent width of the Na I and K I doublets as a 
function of spectral type. As noted in the text, we represent spectral type numerically
with M8$\equiv$-2, L0$\equiv$0 and L5$\equiv$5. The equivalent widths are measured in \AA ngstroms.}
\end{figure}

\begin{figure}
\plotone{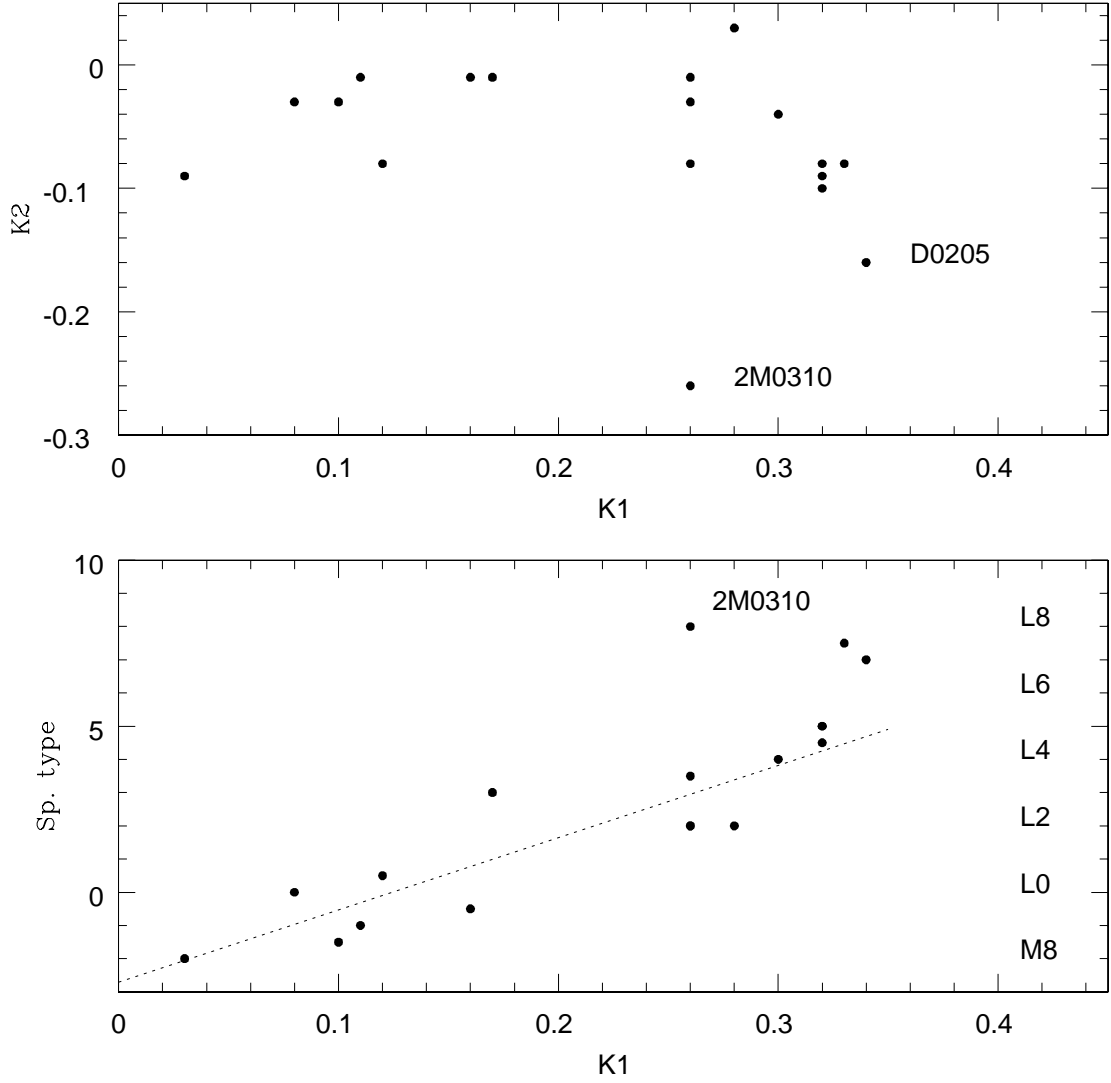}
\caption{ The K1 and K2 indices defined by Tokunaga \& Kobayashi (1999). K2 provides 
a measure of H$_2$ absorption, and the upper panel shows that Denis 0205 and 2M0310 stand
out in the present sample. The lower
panel shows that K1 is well correlated with optical spectral type for dwarfs
earlier than L6; the dotted line shows the linear fit to the data cited in the text.}
\end{figure}

\begin{figure}
\plotone{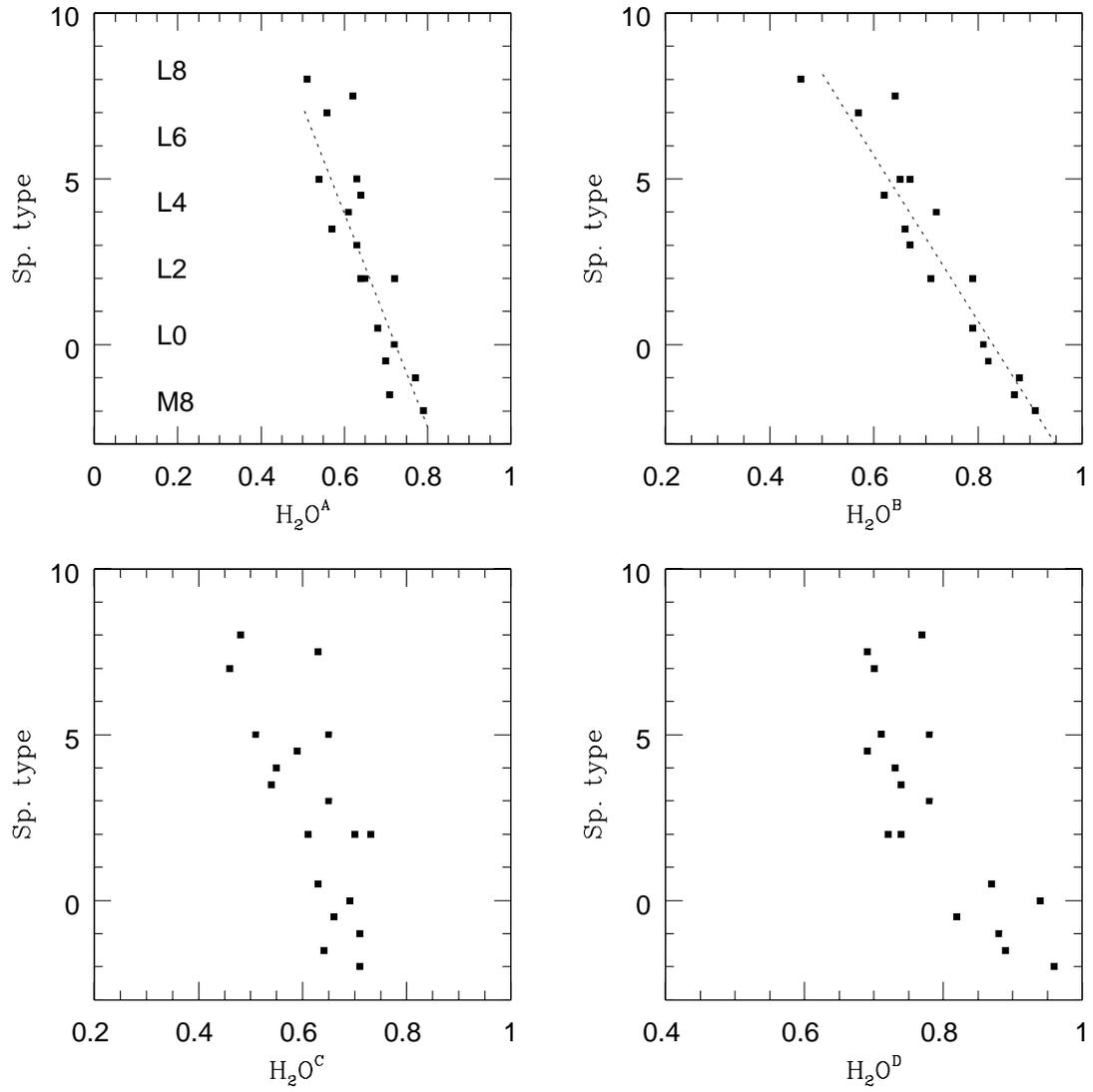}
\caption{ Spectral type as a function of H$_2$O bandstrength. The
dotted lines plotted in the upper two panels are the linear calibrations listed in the text.}
\end{figure}

\end{document}